# Ultra-thin Acoustic Metasurface-Based Schroeder Diffuser

Yifan Zhu[1†], Xudong Fan[1†], Bin Liang[1*], Jianchun Cheng[1*], and Yun Jing[2*]

[1]*Key Laboratory of Modern Acoustics, MOE, Institute of Acoustics, Department of Physics, Collaborative Innovation Center of Advanced Microstructures, Nanjing University, Nanjing 210093, P. R. China*

[2]*Department of Mechanical and Aerospace Engineering, North Carolina State University, Raleigh, North Carolina 27695, USA*

[†]These authors contributed equally to this work.

[*]emails: liangbin@nju.edu.cn (B.L.); jccheng@nju.edu.cn (J.C.C.); yjing2@ncsu.edu (Y.J.)

**Abstract**

"Schroeder diffuser" is a classical design, proposed over 40 years ago, for artificially creating optimal and predictable sound diffuse reflection. It has been widely adopted in architectural acoustics and it has also shown substantial potential in noise control, ultrasound imaging, microparticle manipulation, among others. The conventional Schroeder diffuser, however, has a considerable thickness on the order of one wavelength, severely impeding its applications for low frequency sound. In this paper, a new class of ultra-thin and planar Schroeder diffusers are proposed based on the concept of acoustic metasurface. Both numerical and experimental results demonstrate satisfactory sound diffuse reflection produced from the metasurface-based Schroeder diffuser despite it being one order of magnitude thinner than the conventional one. The proposed design not only offer promising building blocks with great potential to profoundly impact architectural acoustics and related fields, but also constitutes a major step towards real-world applications of acoustic metasurfaces.

In the 1970s, Schroeder published two seminal papers on sound scattering from maximum length sequences and quadratic residue sequences diffusers [1, 2]. For the first time, a simple recipe was proposed to design sound phase grating diffusers with defined acoustic performance. These two papers opened a brand new field of sound diffusers with applications in architectural acoustics [3-5], noise control [6-8], ultrasound imaging [9], microparticle separation [10] and have inspired other disciplines such as energy-harvesting photodiodes [11]. D'Antonio and Konnert [12] presented one of the most accessible review papers examining the theory behind Schroeder's diffusers (SDs). Most importantly, they commercialized SDs and promoted them to be widely adopted in architectural acoustics, where the diffusers can be used to spread the reflections into all directions, reducing the strength of the undesired specular reflection and echo, as well as preserving the sound energy in the space [3]. In contrast to diffusers, sound absorbers reduce the energy in the room, which can be problematic for unamplified performances in concert halls, opera houses, and auditoria. Sound diffusers are also used to promote desired reflections in order to enhance spaciousness in auditoria, to improve speech intelligibility, and to reduce the noise in urban streets [3, 13-14]. Instead of using a surface with random or geometric reflectors, Schroeder innovatively designed a family of diffusers based on number theory sequences, with the ultimate goal to produce predicable and optimal scattering (i.e., the sound is scattered evenly in all directions regardless of the angle of incidence). In spite of the great success that SDs have achieved, they are conventionally designed to have a grating structure with a thickness that is half of the wavelength at the operating frequency in order to achieve the desired phase delays. To put this into perspective, the thickness of an SD reaches a remarkable value of 69 cm at 250 Hz, which is in the range of human voices, truck noises, etc. Figure 1(a) shows a simple one-dimensional (1-D) SD to illustrate the basic concept of SDs. The bulky size of conventional SDs poses a fundamental limitation on their applicability, i.e., SDs are typically limited to mid- and high frequencies because they are too large to be accommodated at low frequencies, which is a very important part of sound that human perceive. In addition, SDs usually do not complement the visual appearance of a space due to their large size and irregular surface. Although active methods may offer a solution to this limitation [15], they are much more expensive and complicated and therefore less practical compared to their passive counterpart.

In this paper, we revisit the SD and redesign it using the concept of acoustic metasurface [16-25]. Despite the considerable efforts dedicated to the research on acoustic metamaterials and acoustic metasurfaces [16-39], they are still at an embryonic stage from the real-world application perspective. Metasurfaces, in particular, are thin structures having subwavelength thickness consisting of unit cells that could give rise to numerous intriguing phenomena such as super sound absorption [16-17], wavefront shaping [18-22], dispersion-free phase engineering [23], and asymmetric acoustic transmission [24-25]. Here we show the potential of using acoustic metasurfaces to break down the fundamental physical barrier in designing ultra-thin SDs. As will be demonstrated in this paper, the metasurface-based SD (MSD) has a comparable performance to the conventional SD that has already been commercialized and widely used in practice. More importantly, the MSD is one order of magnitude thinner with a planar configuration and therefore is more suitable for low frequency applications in architectural acoustics or other related fields. This paper will present the theoretical design, numerical simulation, and experimental demonstration of the ultra-thin MSDs with a thickness that is 1/20 of the center frequency wavelength $\lambda_0$. The unit cell of the proposed MSD is a locally resonant element having a relatively simple geometry and its acoustical response can be engineered flexibly and precisely by adjusting a single geometrical parameter, which enables convenient analytical prediction of its acoustical phase response. The metasurface is designed in a way that the thickness is minimized while the performance is not significantly affected by the viscosity effect [40-41]. This is in contrast to the widely studied space-coiling structure-based metasurfaces which may be liable to suffer from large viscous losses at a comparable thickness [19-21, 23]. Our initial design is further improved by the broadened frequency band introduced by a hybrid structure containing units operating at multiple optimal frequencies. The experimental and simulation results were in good agreement and both showed that the MSD yielded a performance on a par with the conventional SD, despite it being one order of magnitude thinner. This study, for the first time, attempts to bridge the gap between acoustic metasurfaces and their applications to real-world problems.

First we briefly review the conventional design of SDs and elucidate the fundamental limitation of this design. In order to generate diffuse reflection for different incident acoustic waves, the phase shift at the surface of a SD must yield a specific profile such as a special number sequence [42]. Conventionally, the desired phase delay in a SD is achieved by controlling the sound path in a grating structure, resulting in the fact that the maximum depth of individual unit of grating, also referred to as the "well", must reach a half of the wavelength to ensure that the phase changes within a $2\pi$ range. Figure 1(a) shows schematically the 1-D model of a SD formed by a series of wells, which is for generating diffuse reflections in a two-dimensional (2-D) plane and is called a single plane diffuser [42]. To generate diffuse reflections in three-dimensional (3-D) space, one needs to use a 2-D model shown in Fig. 1(b). In the 1-D case, the depths of the wells are dictated by a mathematical number sequence, such as a quadratic residue sequence (QRS) as shown in Fig. 1(a) for which the sequence number for the nth well, $S_n$, is given by [42]:

$$S_n = n^2 \text{Modulo } N \tag{1}$$

where Modulo indicates the least non-negative remainder, $N$ is the number of wells per period. One example of quadratic residue diffusers with $N = 7$ shown in Fig. 1(a) has $S_n = \{0, 1, 4, 2, 2, 4, 1\}$. The depth $h_n$ of the $n$th well is then determined from the sequence $S_n$ using the following equation:

$$h_n = \frac{S_n \lambda_0}{2N}. \tag{2}$$

The phase delay that a SD needs to yield is previously considered unattainable by a simple structure with a deep-subwavelength size. We will revisit this problem from the perspective of acoustic metasurfaces and demonstrate that it is possible to realize such a phase profile by using properly-designed metasurface units at a deep-subwavelength scale in the thickness direction. The schematic diagram of the proposed MSD is illustrated in Fig. 1(c). The ultra-thin MSD is designed to produce the desired scattering fields mimicking those of SDs, via meta-structure units shown in the inset of Fig. 1(c). The width and thickness of the unit are $D = \lambda_0 / 2$ and $\lambda_0 / 20$, respectively. In this study, the neck width of the cavity $w$ is the only tunable parameter for controlling the phase shift of the meta-structure unit. Although the unit cell is Helmholtz resonator (HR)-like, its cavity width and neck width are much larger than those of the classical HRs with respect to $\lambda_0$. Consequently, the well-established analytical theory for classical HRs (e.g., the lumped model) is not valid anymore and must be revisited (please see Supplementary Note 1).

Figure 2(a) shows the simulated and analytically predicted phase response of the meta-structure unit cells for normally incident waves, which provides us the design for a center frequency at $f_0 = 6860 \text{Hz}$. Note that the phase response does depend on the incidence angle and this is discussed in Supplementary Note 2. Finite element analysis software COMSOL 5.0 is used for numerical simulations. The relatively high frequency chosen in this study is merely for the convenience and precision of experimental characterization (The low-frequency performance is difficult to characterize experimentally due to the fact that the requirement on far-field is challenging to fulfill). Our design, however, is readily scalable and could easily be applied to any audible frequency of interest. The simulation results for the design at $f_0 = 343 \text{Hz}$ are shown in Supplementary Note 3 to verify the scalability of our scheme. By adjusting a single parameter $w$, an almost full $2\pi$ control of reflected phase can be achieved as shown in Fig. 2(a). The triangles in Fig. 2(a) mark the parameters of the prototype diffuser based on the simulated result. The seven discrete phases (which correspond to numbers $0-6$ in Fig. 2b) represent phases of $0 - 2\pi \times 6/7$ with a step of $2\pi \times 1/7$. We then design a QRS for a 2-D sample with $N = 7$, and the sequence number $S_{n,m}$ can be expressed as [42]:

$$S_{n,m} = \left(n^2 + m^2\right) \text{Modulo} N, \tag{3}$$

where $n$ and $m$ represent the row and column number of the unit cells. Thus, for generating the same scattering effect as conventional SDs do, the phase response of the MSD unit cells can be expressed as:

$$\phi_{n,m} = \frac{2\pi\left[\left(n^2 + m^2\right) \text{modulo } N\right]}{N}. \tag{4}$$

The corresponding 2-D QRS is shown in Fig. 2(b). This QRS is obtained with indexes n and m starting from 4 (in order to place the zero depth well at the center of the diffuser) in Eq. (3). The photograph of a MSD sample with 2×2 periods (one period is defined as 7×7 unit cells corresponding to one full QRS) is shown in Fig. 2(c). The acoustically rigid material is chosen as Acrylonitrile-Butadiene-Styrene (ABS) plastics with density $\rho$ =1180 kg/m$^3$ and sound speed c = 2700 m/s, which are much larger than those of air, i.e., $\rho_0$ = 1.21 kg/m$^2$ and c$_0$ = 343 m/s. Figure 2(d) show the schematic diagram of the experiment setup, from which the far-field directivity and near-field acoustic pressure distributions can be measured. The acoustic field scanning is accomplished by a measuring system consisting of Brüel&Kjær PULSE Type 3160 and two 0.25-inch-diameter Bruel&Kjær type-4961 microphones.

Figures 3 and 4 show the numerical and experimental results of the MSD sample for normal incidence and 45°-incidence angles, respectively. Figure 3(a) shows the simulated 3-D far-field scattering patterns of the MSD and a referenced flat plate with the same overall size (marked as Plate). The comparisons between the 3-D far-field scattering patterns of the MSD and SD are shown in Supplementary Note 4. Figure 3(b) shows the measured (Upper) and simulated (Lower) near-field scattered acoustic pressure field distributions of the MSD and plate in the x-z plane. Due to the symmetry of MSDs, the acoustic pressure field distributions in the y-z plane is in theory identical to that in the x-z plane. The acoustic energy is scattered into different directions after impinging upon the sample. Numerous side-lobes with similar magnitudes can be observed and diffuse reflection can be effectively realized by the sample. This is more pronounced in Fig. 3(c), which shows the simulated and the measured far-field scattering directivity of the sample (polar response). The reflected fields of the flat plate in Figs. 3(b) and (c) show that the reflected wave is scattered into primarily a single direction, as expected due to specular reflection. The comparison between these two results shows the effectiveness of our MSD sample at the operating frequency. Similarly, the corresponding results at $45°$-incidence angle are shown in Fig. 4 and satisfactory diffuse reflection effect can be also observed for the MSD.

In order to quantitatively evaluate the performance of the MSD, we use a parameter called normalized diffusion coefficient [42-43]:

$$d_n(\theta) = \frac{d(\theta) - d^r(\theta)}{1 - d^r(\theta)}, \tag{5}$$

where $d(\theta)$ and $d^r(\theta)$ are the diffusion coefficients of the sample and the reference flat surface, which can be computed using the equation below [42-43]:

$$d(\theta) = \frac{\left(\sum_{i=1}^{M} 10^{L_i(\theta)/10}\right)^2 - \sum_{i=1}^{M}\left(10^{L_i(\theta)/10}\right)^2}{(M-1)\sum_{i=1}^{M}\left(10^{L_i(\theta)/10}\right)^2}, \tag{6}$$

where $L_i(\theta)$ are a set of sound pressure levels (SPLs) in the polar response, $M$ is the number of receivers and $\theta$ is the angle of incidence. Figures 3(d) and 4(d) show the simulated and measured $d_n(0°)$ and $d_n(45°)$ versus frequency for the MSD and conventional SD at normal incidence and $45°$-incidence angles, respectively. The discrepancies between experimental results and simulation results are possibly due to the background noise, sample fabrication defect, edge scattering, and positioning error of the microphone receivers. The results demonstrate that the MSD has normalized diffusion coefficients comparable with the conventional SD in the vicinity of the center frequency. While the present study uses a period number $2\times 2$, we have performed a series of simulations to investigate the influence of the period number on $d_n(\theta)$ for the SD and MSD, and the results can be found in Supplementary Note 5.

Since the MSD is featured with a sub-wavelength characteristic, the viscosity effect [40-41] could have a non-trivial effect on the performance of the diffuser. We have numerically investigated the effect of viscosity at $f_0 = 6860 \text{Hz}$ in Supplementary Note 6 and found that it does not significantly change the scattering field. In addition, viscosity is well-known to be frequency-dependent and it is expected that viscosity is even more negligible at lower frequencies (e.g., 100-500 Hz), which are what the MSD is truly designed for. Finally, as mentioned earlier, the unit cells are based on unconventional HRs because of their relatively large neck widths (up to $\lambda_0/4$) while conventional HRs have neck widths on the deep-subwavelength scale. This is why the thickness of the metasurface can be minimized without having to suffer from the adverse effects of the viscosity. Indeed, to the best of our knowledge, the thickness of $\lambda_0/20$ is the smallest that has ever been reported for acoustic metasurfaces manipulating transmitted or reflected waves with experimental verification (excluding the acoustic metasurfaces for absorption purposes).

We have demonstrated that MSDs can be designed to achieve efficient acoustic diffuse reflection in the vicinity of the center frequency. This initial design suffers from the relatively narrow bandwidth due to the resonance nature of the unit cell. We will further enhance the MSD by broadening the operating frequency range which is crucial for certain practical applications. A broadband MSD (BMSD) has a hybrid structure comprising components designed for generating the desired phase delay at multiple frequencies. The multi-frequency QRS is shown in Fig. 5(a), in which $A_n$, $B_n$, $C_n$, and $D_n$ represent four different target frequencies and the subscript *n* represents the number in QRS.

In this manner, the staggered units for four operating frequencies lead to the BMSD design that targets different frequencies and yields a $14 \times 14$ array. Figure 5(b) shows the photograph of a BMSD sample. We designed two samples [denoted as BMSD1 and BMSD2 in Figs. 5(c) and 5(e)] with different target frequencies. Figure 5(c) shows the unit parameters of BMSD1 for realizing seven discrete phases ranging from $0 - 2\pi \times 6/7$. Figure 5(d) shows the numerical and experimental results of $d_n(0^\circ)$ and $d_n(45^\circ)$ in the x-z plane versus frequency for the SD and BMSD, respectively. The targeted four frequencies are marked at the coordinate axis of Fig. 5(d), that is, 5772Hz, 6860Hz, 8153Hz, and 11517Hz for BMSD1 (the thickness 0.25cm is unchanged). The targeted frequency for the reference SD is 6860Hz. The corresponding results for BMSD2 are shown in Figs. 5(e) and (f) with the target frequencies being 6860Hz, 8153Hz, 9690Hz and 11517Hz. Again, these results can be scaled to lower frequencies without extra effort.

In order to characterize the broad-band performance of the BMSD samples, we calculate the average normalized diffusion coefficient as:

$$d_n(\theta, f_1, f_M) = \frac{\sum_{i=1}^{M} d_n(\theta, f_i)}{M}, \quad (7)$$

where $d_n(\theta, f_i)$ is the normalized diffusion coefficient at different discrete frequencies of $f_i$, $f_1$ and $f_M$ are the lower and upper bound frequencies of the frequency range of interest, respectively, and $M$ is the number of the simulated discrete frequencies (e.g., we have simulated 13 evenly spaced frequencies within $6292\text{Hz} - 7479\text{Hz}$ for MSD). Comparing SD and MSD, the simulated (measured) $d_n(0^\circ, 6292, 7479)$ are 0.56 and 0.50 (0.35), respectively. The simulated (measured) $d_n(45^\circ, 6292, 7479)$ are 0.35 and 0.37 (0.39), respectively. These results suggest that the MSD have comparable performance as the SD in a relatively small frequency range. In a larger frequency range, $d_n(0^\circ, 6860, 11517)$ (Here, $6860\text{Hz} - 11517\text{Hz}$ covers the targeted frequencies of BMSD1 and BMSD2 with 37 evenly spaced frequencies) for SD, MSD, BMSD1 and BMSD2 are 0.53, 0.15 (0.23), 0.51 (0.28) and 0.49 (0.33), respectively. $d_n(45^\circ, 6860, 11517)$ are 0.38, 0.22 (0.23), 0.34 (0.25) and 0.36 (0.28), respectively. While the SD still has fairly good performance due to the fact that at different frequencies, the phase response at the surface of SD is also a random distribution, the performance of MSD deteriorates dramatically. On the other hand, the performance of the BMSD is comparable to the SD, although it is one order of magnitude thinner. Figure 6 maps the simulated 3-D far-field scattering patterns and the measured and simulated scattered acoustic pressure fields for BMSD1 in the x-z plane for normal incidence and $45^\circ$-incidence angles at 5772Hz, 6860Hz, and 8153Hz, respectively. The experiment results and simulation results are in reasonable agreement. The scattered acoustic fields show that the BMSD yields diffuse reflection at different frequencies. The results suggest that the bandwidth can be broadened by using the BMSD structure and the performance is on a par with the widely commercialized SD.

In conclusion, we have designed an ultra-thin Schroeder diffuser based on the concept of acoustic metasurfaces. The thickness of the sample is only $\lambda_0/20$, exactly one order of magnitude smaller than that of conventional SDs. The proposed diffuser, in theory, can be designed to be even thinner, with the caveat in mind that the viscosity effect will become more dominant and introduce additional absorptions, which could be unwanted in architectural acoustic applications. On the other hand, these additional absorptions may enable hybrid surfaces with simultaneous diffusion and absorption, which can find utility in noise control or in places such as studios where both low reverberation and sound uniformity are desired. We have also proposed a hybrid structure containing units operating at different frequencies in order to broaden the bandwidth of the MSD. The numerical and experimental results both show sound diffuse reflection comparable with the conventional SD. While our study has examined one possible scheme to broaden the bandwidth of the MSD, other feasible schemes will be exploited in the near future under the framework built by this work, including iterative optimization and fractals [42]. Our work takes a first step in applying acoustic metasurfaces to solving practical acoustic problems. The conventional sound diffuser that has been widely adopted in industry is markedly improved by new designs. Our findings may provide a roadmap to manipulate sound scattering and have far-reaching implications in architectural acoustics, noise control and beyond.

**Acknowledgements.**

This work was supported by the National Natural Science Foundation of China (Grant Nos. 11634006 and 81127901) and A Project Funded by the Priority Academic Program Development of Jiangsu Higher Education Institutions.

**Figure Captions**

FIG. 1. (Color online) A one-dimensional Schroeder diffuser (1-D SD). (b) A 2-D SD. (c) The proposed metasurface-based Schroeder diffuser (MSD). The top/bottom images in B and C are the top/ $45^\circ$ angle views of SD and MSD, respectively. The insets in B and C are the cross-sections of unit cells in SD and MSD, with the thicknesses being $\lambda_0/2$ and $\lambda_0/20$, respectively.

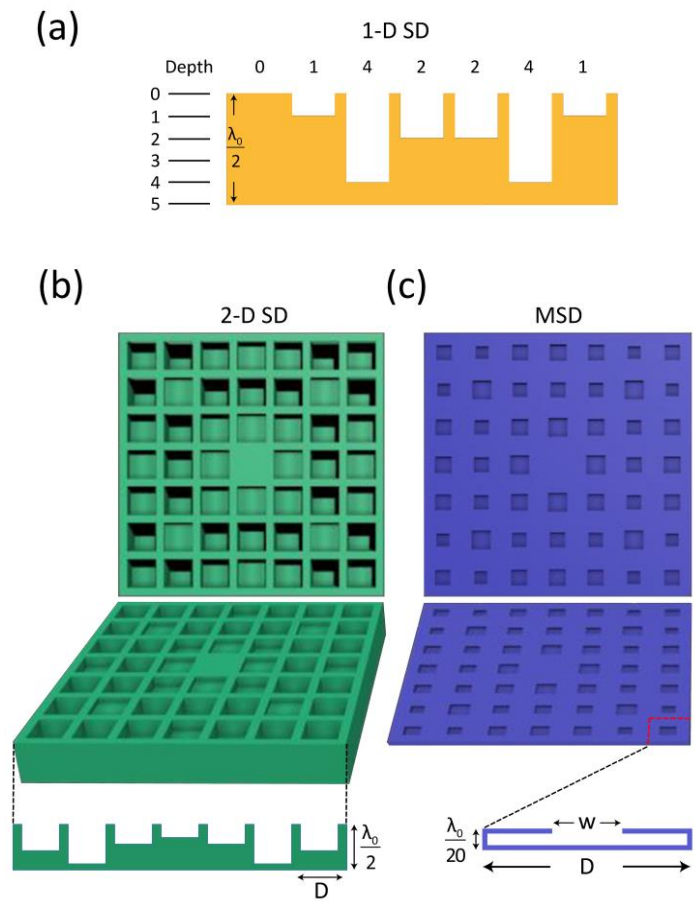

FIG. 2. (Color online) Design of the metasurface-based Schroeder diffuser (MSD). (a) The analytical and simulated relationship between the phase shift and the geometrical parameter $w$ of the MSD at the center frequency of $f_0 = 6860\text{Hz}$. The triangles represent the discrete points for generating the phase of $0 - 2\pi \times 6/7$ with a step of $2\pi \times 1/7$, corresponding to number $0-6$ in (b). (b) The design of a 2-D MSD based on a 2-D quadratic residue sequence (QRS). One period consists of 7×7 unit cells. (c) The photograph of the 3D printed sample with 2×2 periods of QRS, viz., 14×14 unit cells. (d) The schematic diagram of experimental setup.

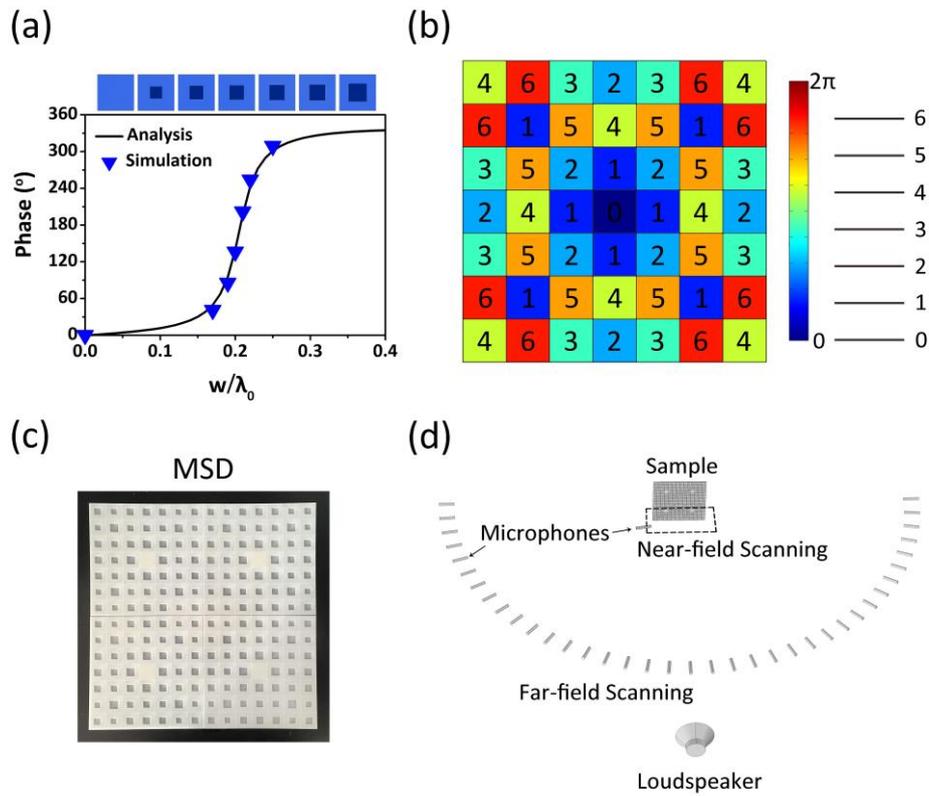

FIG. 3. (Color online) Simulation and experimental results of the MSD for normal incidence. (a) The simulated three-dimensional far-field scattering patterns of the MSD and flat plane with normal incidence. (b) The measured (Upper) and simulated (Lower) scattered acoustic field distributions of the MSD and flat plate in the x-z plane. (c) The simulated and measured scattering field directivity of the MSD (Left) and plate (Right). (d) Simulated and measured normalized diffusion coefficient $d_n(0^o)$ versus frequency for the MSD and SD, respectively.

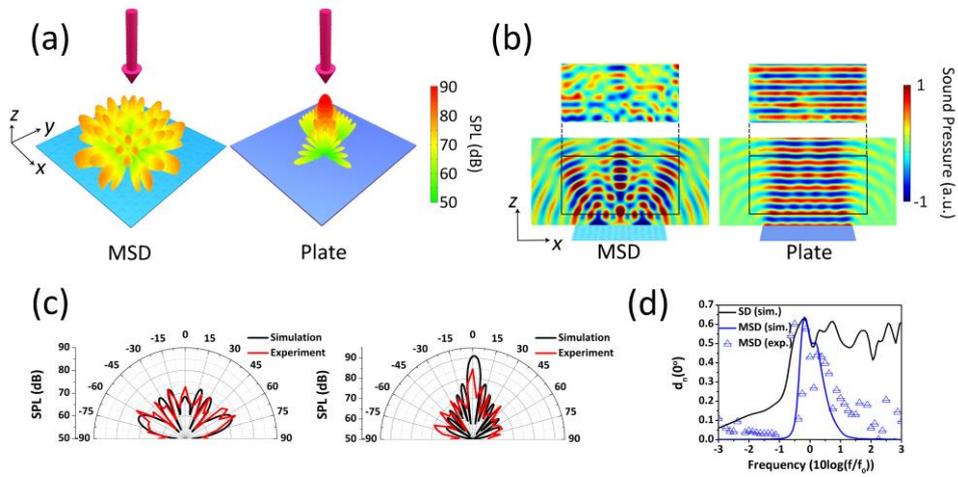

FIG. 4. (Color online) Simulation and experimental results of the MSD for oblique incidence. (a) The simulated three-dimensional far-field scattering patterns of the MSD and flat plane with $45^\circ$-incidence. (b) The measured (Upper) and simulated (Lower) scattered acoustic field distributions of the MSD and flat plate in the x-z plane. (c) The simulated and measured scattering field directivity of the MSD (Left) and plate (Right). (d) Simulated and measured normalized diffusion coefficient $d_n(0^\circ)$ versus frequency for the MSD and SD, respectively.

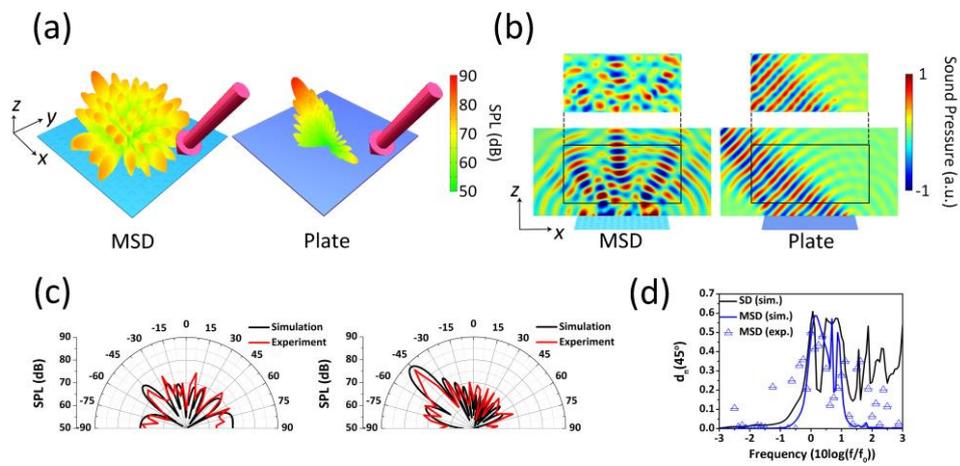

FIG. 5. (Color online) Design of broadband metasurface-based Schroeder diffuser (BMSD). (a) The QRS for a BMSD. $A_n$, $B_n$, $C_n$, and $D_n$ represent four targeted frequencies. (b) Photograph of a 3D printed BMSD sample. (c) The analytical (line) and simulated (triangle) relationship between the reflected phase and the parameter $w$ for four frequency components for BMSD1. (d) Simulated and measured $d_{0,n}$ (Left) and $d_{45,n}$ (Right) in the x-z plane versus frequency for BMSD1 and SD, respectively. The corresponding results for BMSD2 are shown in (e) and (f). In (d) and (f), the center frequency is $f_0 = 6860 \text{Hz}$. The targeted four frequencies of BMSD1 and BMSD2 are marked by squares A, B, C, D in D and F, respectively.

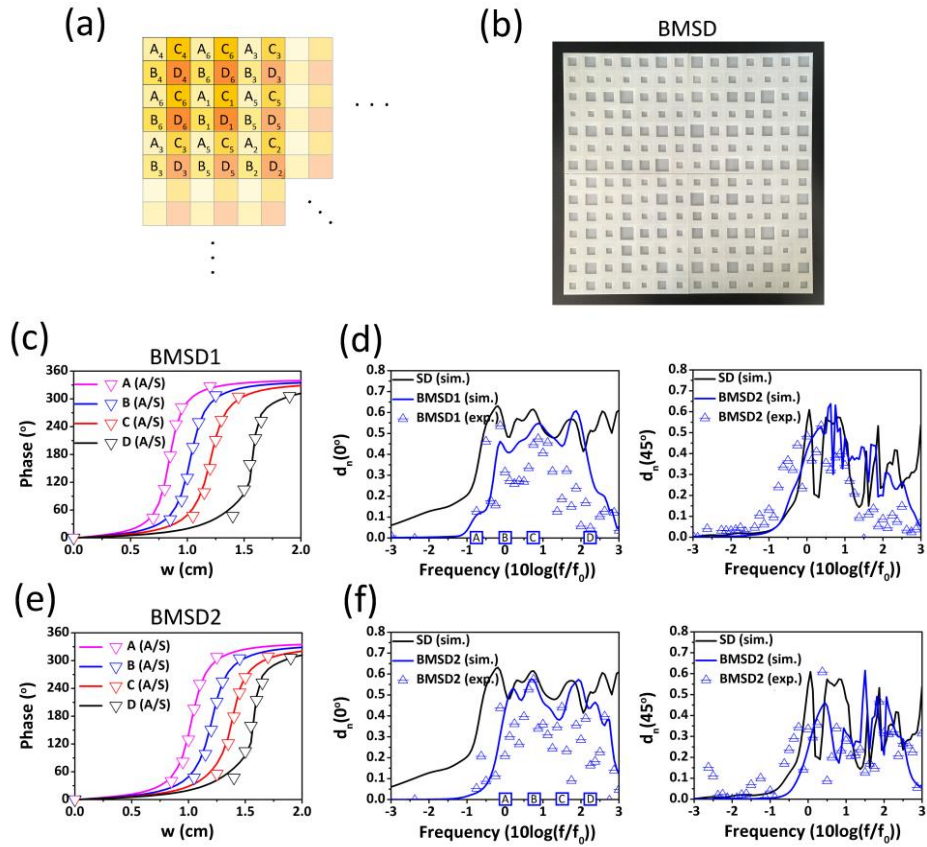
22

FIG. 6. (Color online) Simulation and experiment of BMSD. (a) The simulated three-dimensional far-field scattering patterns of BMSD1 in x-z plane at 5772Hz, 6860Hz, and 8153Hz, respectively, with normal incidence. (b) The measured (Upper) and simulated (Lower) scattered acoustic pressure fields of BMSD1 in x-z plane. The corresponding results for $45^\circ$-incidence angles are shown in (c) and (d).

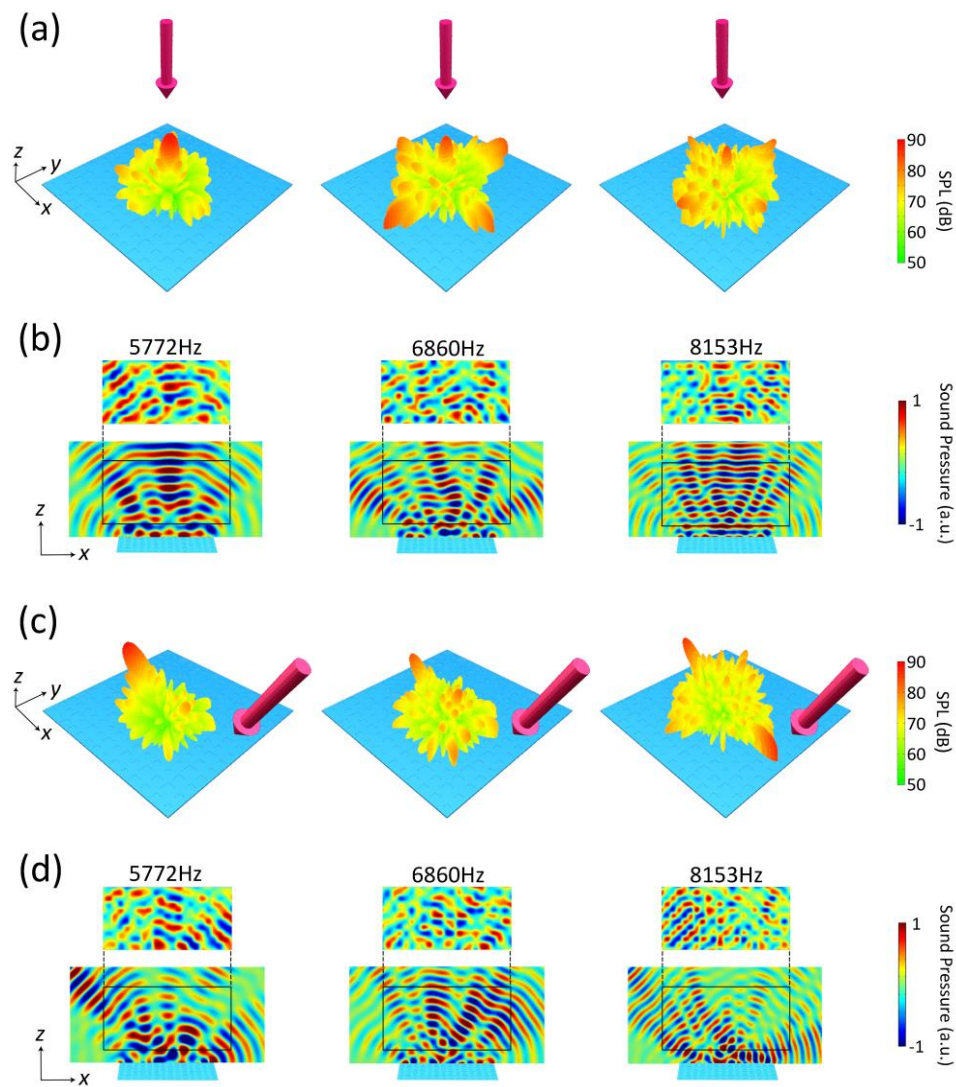